\documentclass[aps,pra,floatfix,showpacs,10pt,twocolumn]{revtex4}
\usepackage{amsfonts}
\usepackage{bbm}
\usepackage{amssymb}
\usepackage[dvips]{graphicx}
\usepackage{amsmath,amsfonts,amssymb,graphics,graphicx,epsfig,color,times,bbm}

\begin{document}

\title{Various correlations in a Heisenberg $XXZ$ spin chain both in thermal equilibrium
and under the intrinsic decoherence}
\author{Jiang-Tao Cai,$^{1}$\footnote{E-mail: jtcai@semi.ac.cn} Ahmad
Abliz,$^{2}$ Shu-Shen Li$^{1}$}

\affiliation{$^{1}$State Key Laboratory for Superlattices and
Microstructures, Institute of Semiconductors, Chinese Academy of
Sciences, P. O. Box 912, Beijing, 100083, P. R. China\\
$^{2}$School of Physics and Electronic Engineering, Xinjiang Normal
University, Urumchi 830054, P. R. China}



\begin{abstract}
In this paper we discuss various correlations measured by the
concurrence (C), classical correlation (CC), quantum discord (QD),
and geometric measure of discord (GMD) in a two-qubit Heisenberg
$XXZ$ spin chain in the presence of external magnetic field and
Dzyaloshinskii-Moriya (DM) anisotropic antisymmetric interaction.
Based on the analytically derived expressions for the correlations
for the cases of thermal equilibrium and the inclusion of intrinsic
decoherence, we discuss and compare the effects of various system
parameters on the correlations in different cases. The results show
that the anisotropy $J_z$ is considerably crucial for the
correlations in thermal equilibrium at zero temperature limit but
ineffective under the consideration of the intrinsic decoherence,
and these quantities decrease as temperature $T$ rises on the whole.
Besides, $J$ turned out to be constructive, but $B$ be detrimental
in the manipulation and control of various quantities both in
thermal equilibrium and under the intrinsic decoherence which can be
avoided by tuning other system parameters, while $D$ is constructive
in thermal equilibrium, but destructive in the case of intrinsic
decoherence in general. In addition, for the initial state
$|\Psi_1(0) \rangle = \frac{1}{\sqrt{2}} (|01 \rangle + |10
\rangle)$, all the correlations except the CC, exhibit a damping
oscillation to a stable value larger than zero following the time,
while for the initial state $|\Psi_2(0) \rangle = \frac{1}{\sqrt{2}}
(|00 \rangle + |11 \rangle)$, all the correlations monotonously
decrease, but CC still remains maximum. Moreover, there is not a
definite ordering of these quantities in thermal equilibrium,
whereas there is a descending order of the CC, C, GMD and QD under
the intrinsic decoherence with a nonnull $B$ when the initial state
is $|\Psi_2(0) \rangle$.
\end{abstract}

\maketitle

\section{introduction}
The most fascinating nonlocal correlation feature of quantum
mechanics is the quantum entanglement, generally considered as an
essential resource for the quantum information processing
\cite{NielsenBook} that provides the possibility of quantum
teleportation \cite{BennettPRL1}, quantum dense coding
\cite{BennettPRL2}, and quantum cryptographic key distribution
\cite{EkertPRL}, etc. Ever since the foundation of the quantum
information science, the quantification of the entangled states has
been one of the most fundamental and substantial tasks. As is well
known, quantum states have been subdivided into entangled states and
separable (nonentangled) states. However, recent research reveals
that entanglement doesn't provide all aspects of quantum
correlations which arise from the noncommutativity of operators
representing states, observables, and measurements \cite{LuoPRA1}.
Quantum states display other nonlocal correlations not present in
the classical counterpart, such as the so-called quantum discord
(QD) that is intimately relevant to local measurement which accounts
for all nonclassical correlations originally introduced by Ollivier
and Zurek \cite{OllivierPRL}. QD is defined by the distinction
between the two quantum extensions of the classical mutual
information defined equivalently in two classical ways and shown to
be nonzero both theoretically \cite{LuoPRA1,OllivierPRL} and
experimentally \cite{LanyonPRL} for some separable states which may
be utilized to speed up some tasks over their classical
counterparts. In addition, QD is responsible for the quantum
computational efficiency of deterministic quantum computation with
one pure qubit in Ref. \cite{DattaPRL,LanyonPRL} albeit in the
absence of entanglement. Therefore, it is imperative and desirable
to study QD aiming at understanding well the relationship between QD
and other correlation indicators and also that among the total
correlations, genuinely classical correlations and purely quantum
ones.

Recently, the QD has been intensively investigated in the literature
both theoretically
\cite{LuoPRA2,SarandyPRA,WerlangPRA1,AliPRA,WerlangPRA2,FanchiniPRA,SunzhaoyuPRA,
LuoPRA3,CilibertiPRA,LiuPRA,LizhenniPRA,LiboPRA,QasimiPRA,ParasharPRA,
GirolamiPRA,DillenschneiderPRB,ShabaniPRL,LangPRL,DakicPRL,StreltsovPRL,YuanJPB,
SunzheJPB,GuojinliangJPB,LuQIC} and experimentally
\cite{LanyonPRL,XuJinshiNatCommun}. Generally, it is somewhat
difficult to calculate QD and the analytical solutions can hardly be
obtained except for some particular cases, such as the so-called $X$
states \cite{AliPRA}. Some researches show that QD, concurrence (C)
and classical correlation (CC) are respectively independent measures
of correlations with no simple relative ordering and QD is more
practical than entanglement \cite{LanyonPRL}. Quite rencently,
Daki\'{c} $et$ $al$ \cite{DakicPRL} have introduced an easily
analytically computable quantity, geometric measure of discord
(GMD), and given a necessary and sufficient condition for the
existence of nonzero QD for any dimensional bipartite states.
Moreover, the dynamical behavior of QD in terms of decoherence
\cite{MazieroPRA1,MazieroPRA2,MazzolaPRL,YuanJPB,LuQIC} in both
Markovian \cite{WerlangPRA1} and Non-Markovian
\cite{FanchiniPRA,WangPRA,VasilePRA} cases is also taken into
account.

In previous studies, the influence of intrinsic (phase) decoherence,
a virtually unavoidable effect caused by the interaction of the
system with the surrounding environment, on the dynamics of various
correlations (C, QD, GMD and CC) using a Heisenberg spin chain as a
quantum channel has not been considered. Also the effect of
Dzyaloshinskii-Moriya (DM) interaction, which is introduced via the
extension of the Anderson superexchange interaction theory by
including the spin-orbit coupling effect, on these correlations and
the comparison of these quantities have been rarely reported in the
literature. On the other hand, due to the good integrability and
scalability, the solid state systems
\cite{NielsenPRA,WangxiaoguangPRA,KamtaPRL,O'ConnorPRA,SunPRA,YeoPRA,AhmadJPB,
CaiOpt} have gained great attention. Particularly, the Heisenberg
spin chains, as the natural candidates for the realization of the
entanglement showed some substantial advantages compared with the
other physical systems
\cite{LossPRA,LossPRB,KaneNature,SorensenNature,LiuwumingPRL}. In
addition, by suitable coding, the Heisenberg interaction alone can
be used for quantum computation
\cite{LidarPRL,DivincenzoNature,SantosPRA}. To this end, in this
paper we investigate in detail both the thermal equilibrium and
dynamical behaviors of various correlations in a Heisenberg $XXZ$
spin-$\frac{1}{2}$ chain. We find that the anisotropy $J_z$ is
considerably crucial for these quantities in thermal equilibrium at
zero temperature limit but ineffective under the consideration of
the intrinsic decoherence, and these quantities decrease as
temperature $T$ rises on the whole. Besides, $J$ and $D$ contribute
equivalently to these quantities and turn out to be the most
efficient controlling parameters. $J$ plays a constructive role and
$B$ a detrimental role in the manipulation and control of these
quantities both in thermal equilibrium and under the consideration
of intrinsic decoherence, while $D$ is constructive in thermal
equilibrium, but becomes to be destructive in the decoherent time
evolution process in general. Albeit $B$ is detrimental for these
quantities, it remains worthy of being studied since the inclusion
of it is in practical need such as the nuclear magnetic resonance
quantum computing and the superconducting quantum computing.
Furthermore, when the initial state is $|\Psi_1(0) \rangle =
\frac{1}{\sqrt{2}} (|01 \rangle + |10 \rangle)$, all the
correlations except the CC, exhibit a damping oscillation to a
stable value larger than zero following the time, while for the
initial state $|\Psi_2(0) \rangle = \frac{1}{\sqrt{2}} (|00 \rangle
+ |11 \rangle)$, all the correlations monotonously decrease, but CC
still remains maximum. Moreover, there is not a definite ordering of
these quantities in thermal equilibrium, whereas there is a
descending order of the CC, C, GMD and QD under the intrinsic
decoherence with a non-zero $B$ when the initial state is
$|\Psi_2(0) \rangle$.

This paper is organized as follows. In section II, we present a
brief overview of various correlation measured quantities. Next we
study the thermal correlations in a two-qubit Heisenberg $XXZ$ spin
chain with DM interaction in the presence of an external magnetic
field along the $z$-axis in section III. Subsequently, we turn to
the influence of intrinsic decoherence on various quantities in
section IV. Finally we conclude the paper in section V.

\section{Correlation measures for bipartite system}
Firstly we give a brief overview of various correlation measures.
Given a bipartite quantum state $\rho_{AB}$ in a composite Hilbert
space $\mathcal {H}=\mathcal {H}_{A}\otimes\mathcal {H}_{B}$, the
concurrence \cite{WoottersPRL} as an indicator for entanglement
between the two qubits is
\begin{eqnarray}
\label{eq:1}
 C(\rho_{AB})=\max\{\lambda_1-\lambda_2-\lambda_3-\lambda_4,0\},
\end{eqnarray}
where $\lambda_i(i=1,2,3,4)$ are the square roots of the eigenvalues
of the "spin-flipped" density operator
$R=\rho\widetilde\rho=\rho(\sigma^y_1\otimes\sigma^y_2)\rho^*(\sigma^y_1\otimes\sigma^y_2)$
in descending order. $\sigma^y$ is the Pauli matrix and $\rho^*$
denotes the complex conjugation of the matrix $\rho$ in the standard
basis $\{|00\rangle, |01\rangle, |10\rangle,|11\rangle\}.$ For the
density matrix in the $X$ form, an alternative equivalent expression
is given by
\begin{eqnarray}
\label{eq:2}
 C(\rho_{AB})=\max\{C_1,C_2,0\},
\end{eqnarray}
where $C_1 = 2(|\rho_{41}| - \sqrt{\rho_{33}\rho_{22}})$ and $C_2 =
2(|\rho_{32}| - \sqrt{\rho_{44}\rho_{11}}).$

In classical information theory \cite{CoverBook}, the total amount
of correlations between two systems \textbf{A} and \textbf{B} can be
represented by the classical mutual information
$I(A;B)=H(A)+H(B)-H(A,B),$ where $H(A)\big(H(B),H(A,B)\big)$ is the
Shannon entropy $H=-\sum_{i}p_{i}\log_{2}p_{i},$ $p_{i}$
representing the probability of an event $i$ associated with
\textbf{A} (\textbf{B}, \textbf{AB}). By virtue of the Bayes rule,
the mutual information can be rewritten as $J(A;B)=H(A)-H(A|B),$ in
which $H(A|B)=H(A,B)-H(B)$ is the classical conditional entropy
employed to quantify the ignorance of the state of \textbf{A} when
one knows the state of \textbf{B}. Despite the equivalency of the
two expressions in the classical case, the quantum versions of the
two are not equivalent anymore. In the generalized quantum version,
the classical probability distributions are replaced by the density
operator $\rho$ and the Shannon entropy by the von Neumann entropy
\cite{WehrlRevModPhys}
$S(\rho)$=$-\textrm{Tr}$($\rho$$\log_{2}\rho$). Accordingly, the
quantum version of the two mutual information expressions can be
obtained as
\begin{eqnarray}
\label{eq:3}
 \mathcal{I}(\rho_{A};\rho_{B})=S(\rho_{A})+S(\rho_{B})-S(\rho_{AB}),
\end{eqnarray}
where $\rho_{A(B)}=\textrm{Tr}_{B(A)}(\rho_{AB})$ is the reduced
density matrix of the subsystem \textbf{A}(\textbf{B}) by tracing
out the subsystem \textbf{B}(\textbf{A}). The quantum generalization
of the conditional entropy is not the simply replacement of Shannon
entropy with von Neumann entropy, but through the process of
projective measurement on the subsystem \textbf{B} by a set of
complete projectors ${B_k},$ with the outcomes labeled by $k,$ then
the conditional density matrix $\rho_k$ becomes
\begin{eqnarray}
\label{eq:4}
 \rho_{k}=\frac{1}{p_k}(\mathbbm{l}_A \otimes {B_k})\rho(\mathbbm{l}_A \otimes {B_k}),
\end{eqnarray}
which is the locally post-measurement state of the subsystem
\textbf{B} after obtaining the outcome $k$ on the subsystem
\textbf{A} with the probability
\begin{eqnarray}
\label{eq:5}
 p_k=\textrm{Tr} [(\mathbbm{l}_A\otimes{B_k})\rho(\mathbbm{l}_A\otimes{B_k})],
\end{eqnarray}
where $\mathbbm{l}_A$ is the identity operator on the subsystem
\textbf{A}. The projectors ${B_k}$ can be parameterized as
${B_k}=V|k\rangle\langle k|V^{\dagger},k=0,1$ and the transform
matrix $V\in U(2)$ \cite{SarandyPRA} is
\begin{equation}
\label{eq:6} V=\left(
\begin{array}{cccc}
    \cos\theta &   e^{-i\phi} \sin\theta \\
    e^{i\phi} \sin\theta & -\cos\theta  \\
\end{array}
\right).
\end{equation}
Then the conditional von Neumann entropy (quantum conditional
entropy) and quantum extension of the mutual information can be
defined as \cite{OllivierPRL}
\begin{eqnarray}
\label{eq:7}
 S(\rho|\{B_k\})=\sum_{k} p_k S(\rho_k),
\end{eqnarray}
\begin{eqnarray}
\label{eq:8}
 \mathcal{J}(\rho_{AB}|\{B_k\})=S(\rho_{A})-S(\rho|\{B_k\}).
\end{eqnarray}
Following the definition of the CC in Ref.\cite{OllivierPRL}
\begin{eqnarray}
\label{eq:9}
 \mathcal {CC}(\rho_{AB})=\sup_{\{B_k\}} \mathcal{J}(\rho_{AB}|\{B_k\}),
\end{eqnarray}
then QD defined by the difference between the quantum mutual
information $\mathcal{I}(\rho_{AB})$ and the $\mathcal
{CC}(\rho_{AB})$ is given by
\begin{eqnarray}
\label{eq:10}
 \mathcal {QD}(\rho_{AB})=\mathcal{I}(\rho_{AB}) - \mathcal{CC}(\rho_{AB}).
\end{eqnarray}
If we denote $S_{\min} (\rho_{AB}) = \min_{\{B_k\}}
S(\rho_{AB}|\{B_k\}),$ then a variant expression of QD reads
\cite{SunzhaoyuPRA}
\begin{eqnarray}
\label{eq:11}
 \mathcal {QD}(\rho_{AB})=S(\rho_{B})-S(\rho_{AB})+S_{\min} (\rho_{AB}).
\end{eqnarray}

It is usually difficult to get the analytical expression of QD
except for some special cases, thus another correlation measure,
GMD, is introduced by Daki\'{c} $et$ $al$ \cite{DakicPRL} to
simplify the computation. It is defined as
\begin{eqnarray}
\label{eq:12}
 \mathcal {D}_{G}(\rho_{AB})=\min_{\chi} \| \rho-\chi \|^{2},
\end{eqnarray}
where the minimum is over the set of zero-discord states and the
geometric quantity
\begin{eqnarray*}
\label{eq:11}
 \| \rho-\chi \|^{2} := \textrm{Tr} (\rho-\chi) ^{2}
\end{eqnarray*}
is the square of Hilbert-Schmidt norm of Hermitian operators. For
any two-qubit state in the so-called Bloch basis
\begin{eqnarray*}
\label{eq:13}
 \hspace*{5mm}\rho=\frac{1}{4} \sum_{i,j=0}^{3} R_{ij} \sigma_i \otimes \sigma_j
     =\frac{1}{4} \bigg(\mathbbm{l}_{A} \otimes \mathbbm{l}_{B}+\sum_{i=1}^{3}(x_i
     \sigma_i \otimes \mathbbm{l}_{B}\qquad
\end{eqnarray*}
\vspace*{-5mm}
\begin{eqnarray}
\qquad\qquad\hspace*{-24mm}+ y_i \mathbbm{l}_{B} \otimes \sigma_i)
     + \sum_{i,j=1}^{3} t_{ij} \sigma_i \otimes \sigma_j\bigg),
\end{eqnarray}
where $R_{ij} = \textrm{Tr}[\rho(\sigma_i \otimes \sigma_j)],$
$\sigma_0 = \mathbbm{l}_{2\times2},$ $\sigma_i (i = 1,2,3)$ are the
Pauli matrices, $\vec{x}=\{ x_i \}$, $\vec{y}=\{ y_i \}$ are the
three-dimensional Bloch vectors associated with subsystems
\textbf{A}, \textbf{B}, and $t_{ij}$ denote the elements in the
correlation matrix $T$. Then, a variant expression of GMD is given
by
\begin{eqnarray}
\label{eq:14}
 \mathcal {D}_{G}(\rho_{AB})=\frac{1}{4}(\| \vec{y} \vec{y}^{\textrm{T}}
  \| + \| T \|^{2} - k_{\max}),
\end{eqnarray}
where $k_{\max}$ is the largest eigenvalue of the matrix $\vec{y}
\vec{y}^{\textrm{T}} + T^{\textrm{T}}T$ (in the case of measurement
on the subsystem \textbf{A}, one needs to replace $\vec{y}$ with
$\vec{x}$ and $T^{\textrm{T}}T$ with $TT^{\textrm{T}}$). An
alternative formulation for GMD has been provided in \cite{LuoPRA3}.
Note that its maximum value is $\frac{1}{2}$ for two-qubit states,
so it is appropriate to consider 2$D_{G}$ as a measure of GMD
hereafter in order to compare with other correlation measures
\cite{GirolamiPRA}.

\section{The correlations in a Heisenberg $XXZ$ spin chain in thermal equilibrium}
The physical system we discuss here is a two-qubit Heisenberg $XXZ$
spin chain with the DM anisotropic antisymmetric interaction
\cite{DzyaloshinskiiJPhysChemSolids,MoriyaPRL} under the external
magnetic field and its Hamiltonian is written as
\begin{eqnarray*}
\label{eq:15}\hspace*{9mm} H=\frac{1}{2}
[J\left(\sigma_{1}^{x}\sigma^{x}_{2}+\sigma_{1}^{y}\sigma^{y}_{2}\right)+
J_{z}\sigma_{1}^{z}\sigma_{2}^{z}+B\left(\sigma_{1}^{z}+\sigma_{2}^{z}\right)\qquad
\end{eqnarray*}
\vspace*{-7mm}
\begin{eqnarray}
\qquad\qquad\hspace*{-30mm}+D\left(\sigma_{1}^{x}\sigma_{2}^{y}-\sigma_{1}^{y}\sigma_{2}^{x}\right)],
\end{eqnarray}
where $J$ and $J_z$ are the real coupling constants,
$\sigma_{i}^{j}(i=1,2;j = x, y, z)$ are the Pauli matrices. $B$ and
$D$ are respectively the $z$-component of the external magnetic
field and DM interaction. We are working in units, so that all
parameters are dimensionless. The state of a typical solid-state
system at thermal equilibrium in temperature $T$ (canonical
ensemble) is $\rho(T)= e^{- \frac{H}{kT} } / Z,$ with
$Z=\textrm{Tr}(e^{-\frac{H}{kT}})$ the partition function and $k$
the Boltzmann constant. Usually we work with natural unit system
$\hbar = k = 1$ for simplicity and henceforth.

In the first instance, we derive the analytical expressions for
various correlation measured quantities. After some straightforward
algebra, one can readily obtain the thermal concurrence
\begin{eqnarray}
\label{eq:16}
 C(\rho_{AB})=\max\{\frac{1}{Z}\big(2 e^{\frac{J_z}{2T}} \sinh\frac{\mu}{T} - e^{-\frac{J_z}{2T}}\big),0\},
\end{eqnarray}
where $\mu=\sqrt{J^{2}+D^{2}}$ and $Z = e^{-\frac{2B+J_z}{2T}}
\big(1 + e^{\frac{2B}{T}} + 2e^{\frac{B+J_z}{T}}
\cosh\frac{\mu}{T}\big) $. Next, in order to gain the thermal CC and
thermal QD, one needs to calculate the quantum conditional entropy
and minimize it over all possible projective measurements which is
the most difficult part. As per Eqs. (\ref{eq:4})-(\ref{eq:7}),
after the minimization of the quantum conditional entropy (i.e., to
set the derivative of the quantum conditional entropy
$S(\rho|\{B_k\})$ with respect to angels $\theta$ and $\phi$ to be
zero), one can find that the quantum conditional entropy is
independent of angle $\phi$, and reaches its minimum value when
$\theta = \frac{(2m+1)\pi}{4}$ (m $\in \mathbb{Z}$) for $-J < J_z <
J$ in the absence of $B$ and $D$. It is independent of $\theta$ when
$J_z = \pm J$ (i.e., the $XXX$ model, this can be explained from the
physical respective as the system is isotropic). Otherwise its
minimum value is reached as $\theta = \frac{m \pi}{2}$ (m $\in
\mathbb{Z}$). As $B$ and $D$ are introduced, the range of $J_z$ ($-J
< J_z < J)$, in which the quantum conditional entropy reaches its
minimum value when $\theta = \frac{(2m+1) \pi}{4}$ (m $\in
\mathbb{Z}$), are broadened. Moreover, the effect of $B$ on the
dependence of the minimum quantum conditional entropy with respect
to angel $\theta$ is different for the antiferromagnetic (AFM) case
($J>0$) and ferromagetic (FM) case ($J<0$) with $D$ fixed. The
effect for the AFM case is more significant compared with the case
of FM, in which the range of $J_z$ ($-J < J_z < J)$ widens slightly.
While the effect of $D$ on the dependence mentioned above are the
same for both the AFM and FM cases when $B$ is fixed. In addition,
the range reduces with the rise of the temperature. Thereby, the CC
and QD are obtained respectively as
\begin{eqnarray}
\label{eq:17}
 \mathcal {CC}(\rho_{AB}) = S(\rho_{A}) - \min \{ \Lambda_{1},\Lambda_{2} \},
\end{eqnarray}

\begin{eqnarray}
\label{eq:18}
 \mathcal {QD}(\rho_{AB}) = S(\rho_{B}) - S(\rho_{AB}) + \min \{ \Lambda_{1},\Lambda_{2} \},
\end{eqnarray}
where
\begin{eqnarray*}
\label{eq}
 S(\rho_{A})= S(\rho_{B})= - \lambda_{-} \log_{2}{\lambda_{-}} -
\lambda_{+} \log_{2}{\lambda_{+}},
\end{eqnarray*}
\vspace*{-5mm}
\begin{eqnarray*}
\lambda_{\pm} = \frac{1}{Z} (e^{- \frac{J_z \pm 2B}{2T}} +
e^{\frac{J_z}{2T}} \cosh \frac{\mu}{T}),
\end{eqnarray*}
\vspace*{-5mm}
\begin{eqnarray*}
S(\rho_{AB}) = - \sum_{\eta_{i}=1}^{4} \eta_{i} \log_{2}{\eta_{i}},
\end{eqnarray*}
\vspace*{-5mm}
\begin{eqnarray*}
\eta_{1,2} = \frac{1}{Z} e^{- \frac{J_z \pm 2B}{2T}},
\eta_{3,4} =
\frac{1}{Z} e^{\frac{J_z \pm 2\mu}{2T}},
\end{eqnarray*}
\vspace*{-5mm}
\begin{eqnarray*}
\Lambda_{1} = \frac{1}{\ln 4} [\ln 4 - (1-\delta) \ln (1-\delta) -
(1+\delta) \ln (1+\delta)],
\end{eqnarray*}
\vspace*{-5mm}
\begin{eqnarray*}
\delta = \frac{2\sqrt{(e^\frac{2B}{T}-1)^2 + 4e^{\frac{2(B+J_z)}{T}}
\sinh^{2} \frac{\mu}{T}}}{1 + e^{\frac{2B}{T}} +
2e^{\frac{B+J_z}{T}} \cosh\frac{\mu}{T}},
\end{eqnarray*}
\vspace*{-5mm} and
\begin{eqnarray*} \Lambda_{2} = (\eta_{1} +
\omega) (\xi_{-} +\xi_{+}) + (\eta_{2} + \omega) (\zeta_{-}
+\zeta_{+}),
\end{eqnarray*}
\vspace*{-5mm}
\begin{eqnarray*}
\xi_{\pm} = -\frac{1 \pm \nu_{1}}{\ln4} \ln(\frac{1 \pm
\nu_{1}}{2}), \zeta_{\pm} = -\frac{1 \pm \nu_{2}}{\ln4} \ln(\frac{1
\pm \nu_{2}}{2}),
\end{eqnarray*}
\vspace*{-5mm}
\begin{eqnarray*}
\omega= \frac{1}{Z} e^{\frac{J_z}{2T}} \cosh \frac{\mu}{T},
\nu_{1,2} = \frac{\sqrt{(e^{\frac{J_z+B}{T}} \cosh
\frac{\mu}{T}-1)^{2}}}{1+ e^{\frac{J_z\pm B}{T}} \cosh
\frac{\mu}{T}}.
\end{eqnarray*}
Finally, according to Eqs. (\ref{eq:13}) and (\ref{eq:14}), thermal
GMD can be written as
\begin{eqnarray}
\label{eq:19}
 2\mathcal {D}_{G}(\rho_{AB}) =  \Omega -
 \frac{1}{2} \max \{ \Gamma_{1},\Gamma_{2} \},
\end{eqnarray}
in which
\begin{eqnarray*}
\Omega = \frac{2 e^{-\frac{J_z}{T}} \cosh \frac{2B}{T} - 4 \cosh
\frac{B}{T}
 \cosh \frac{\mu}{T} + e^{\frac{J_z}{T}} (3\cosh \frac{2 \mu}{T} - 1)}{Z^2},
\end{eqnarray*}
\vspace*{-5mm}
\begin{eqnarray*}
\Gamma_{1} = \frac{4 e^{-\frac{J_z}{T}} [(\cosh \frac{B}{T}
 - e^{\frac{J_z}{T}} \cosh \frac{\mu}{T})^{2} + \sinh^{2} \frac{B}{T}]}{Z^2},
\end{eqnarray*}
\vspace*{-5mm}
\begin{eqnarray*}
\Gamma_{2} = \frac{4 e^{\frac{J_z}{T}} \sinh^{2}
\frac{\mu}{T}}{Z^2}.
\end{eqnarray*}

\begin{figure}[tbp]
\centering
\includegraphics[height=6cm,width=8.5cm]{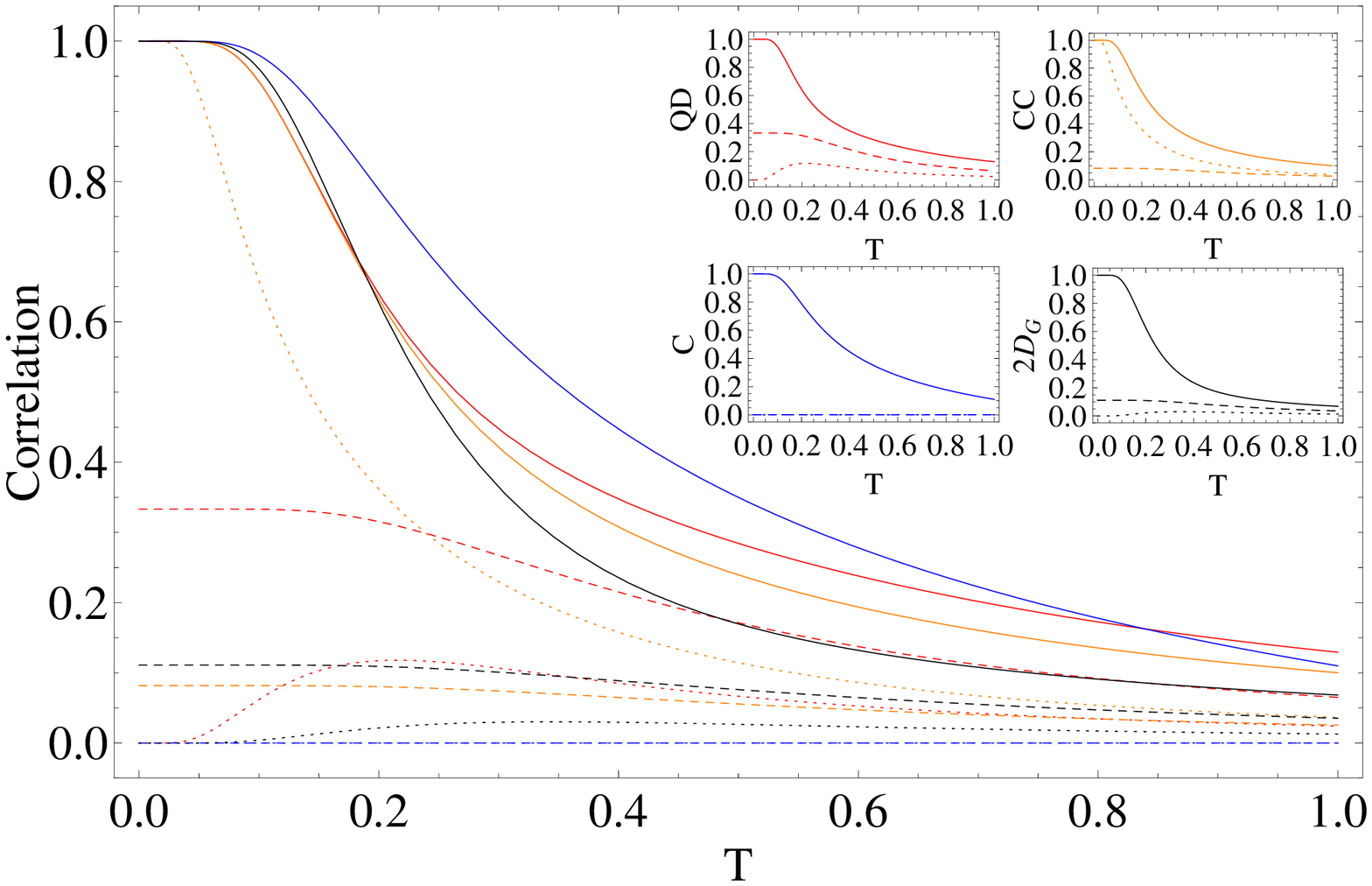}
\caption{(Color online) Various correlations are plotted as a
function of temperature $T$ for different isotropy $J$ with
$J_z=-0.5$ in the absence of $B$ and $D$, in which $J=0.3,0.5,1$
corresponds to dotted, dashed and solid line respectively. QD, CC, C
and GMD corresponds respecitvely to the red, orange, blue and black
line. The inset is the uncombined plot.} \label{fig.1}
\end{figure}

In the second instance, we concentrate on the numerical analysis of
the dependence of various correlations on the different tunable
system parameters at length. Figure \ref{fig.1} plots the behavior
of various quantities versus temperature $T$ for different isotropy
$J$ in the absence of external magnetic field and DM interaction
($B=D=0$). These quantities are invariant under the substitutions $J
\rightarrow -J$ and $D \rightarrow -D$ as well as being contributed
equally by $J$ and $D$ in thermal equilibrium since $J$ and $D$ only
appear in the term $\mu = \sqrt{J^2 + D^2}$. Without loss of
generality, we restrict our attention to the case of $J > 0$. The
figure clearly shows that, when $J < |J_z|$, QD begins at zero and
increases to a certain value as $T$ rises, then decreases with the
further rise of $T$ until reaching the critical temperature, at
which QD vanishes. In addition, this phenomenon can only occur for
the case of an appropriate $J_z$ in the negative region (i.e.
$J_z<0$). This is also true for GMD but not for C and CC. Note that
C is always zero in this case, while CC starts at maximum and
decreases with $T$. When $J \geqslant |J_z|$, QD starts at a
definite value (for $J = |J_z|$) and at the maximum value (for $J
> |J_z|$), then decreases with $T$, which is also valid for GMD and
C except that C is still zero when $J = |J_z|$. However, CC
decreases to a certain value when $J = |J_z|$ and increases
immediately to maximum for $J > |J_z|$ for low temperatures in the
vicinity of zero. The results reveal that the quantum phase
transition (QPT) occurs at $J = |J_z|$. We should also note that QD
is always larger than GMD in this case. But there is no definite
ordering of these measures that are dependent on various system
parameters. Moreover, the critical temperature can be elevated by
the larger absolute value of the isotropy parameter $J$.

\begin{figure}[tbp]
\centering
\includegraphics[height=5.5cm,width=8cm]{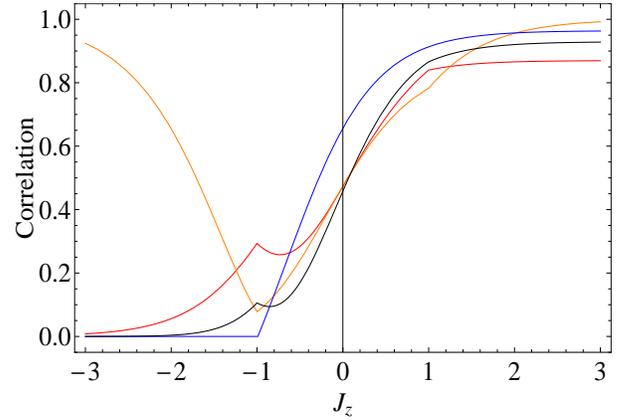}
\caption{(Color online) Various correlations versus the anisotropy
$J_z$ in the absence of B and D, with $J=1$, $T=0.5$. The red,
orange, blue and black line indicates QD, CC, C and GMD
respectively.}\label{fig.2}
\end{figure}

Subsequently, in order to demonstrate the effect of $J_z$ on various
quantities, we plot Fig. \ref{fig.2}, from which one can see that
both QD and GMD are zero when $J_z$ is small in the negative region,
and start to increase with the increasing of $J_z$ at the critical
point until $J_z=-J=-1$, at which they undergo the so-called sudden
change. With the further increase of $J_z$, they decrease slightly
and then continue to increase until reaches two different stable
values (near the maximum). Note that the sudden change also occurs
at $J_z=J=1$ for both of them. However the above process is not true
for C and CC. C is always null in the region $J_z<-1$ and begins to
increase abruptly at $J_z=-1$ until to a stable value as $J_z$
increases and the sudden change does not happen for C. As for CC, it
decreases from the maximum value to the minimum value as $J_z$
increases until $J_z=-1$ and undergoes the sudden change twice at
$J_z=\pm J$, then finally revives to the maximum value with the
further increase of $J_z$. Thereby, by comparing with the result in
Ref. \cite{WerlangPRA2} that QD can signal a QPT at finite
temperature while C can't, we can conclude that not only QD (or
GMD), but also CC can detect the critical points of QPT.

\begin{figure}[tbp]
\centering
\includegraphics[height=6cm,width=8.5cm]{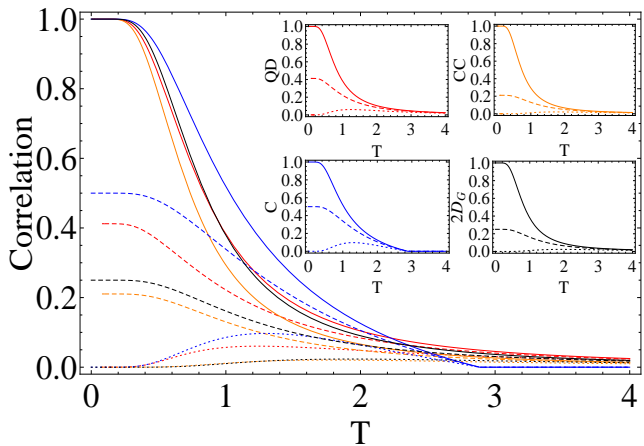}
\caption{(Color online) Various correlations are plotted as a
function of temperature $T$ for different magnetic field $B$ with
$J_z=1$, $D=0$, where $B=0,2,4$ corresponds respectively to the
solid, dashed and dotted line. The red, orange, blue and black line
indicates QD, CC, C and GMD respectively. The inset is the
uncombined plot.}\label{fig.3}
\end{figure}

\begin{figure}[tbp]
\centering
\includegraphics[height=5.5cm,width=8cm]{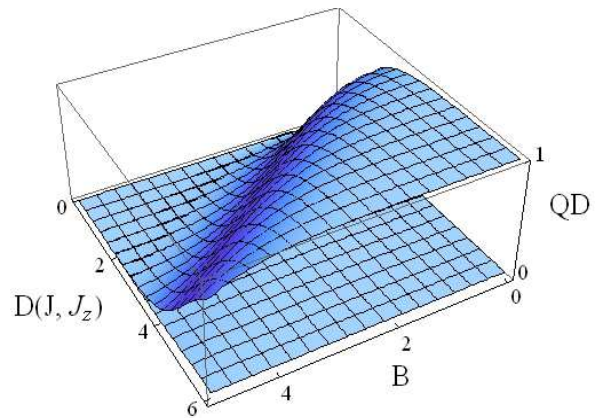}
\caption{(Color online) The QD are plotted as a function of $B$ and
$D$ (or $B$ and $J,$ or $B$ and $J_z$) at temperature $T=0.5$ with
other parameters fixed at the value 0.1. The upper surface plot
represents two completely overlapped surfaces QD($B,D$) and
QD($B,J$), and the lower one corresponds to
QD($B,J_z$).}\label{fig.4}
\end{figure}

\begin{figure}[tbp]
\centering
\includegraphics[height=6cm,width=8.5cm]{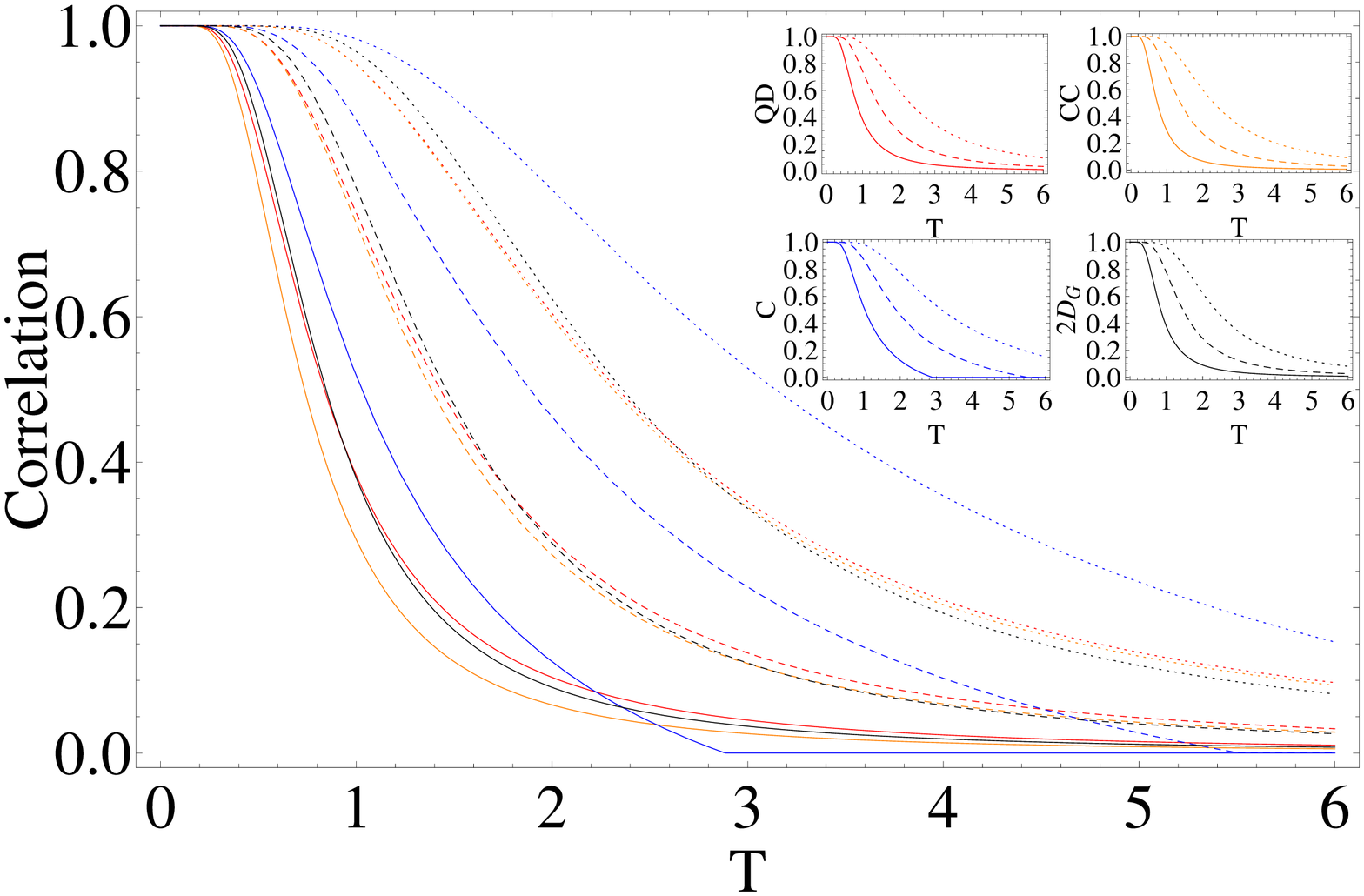}
\caption{(Color online) Various correlations are plotted as a
function of temperature $T$ for different $D$ with $J_z=1$, $B=0$,
where $D=0,2,4$ corresponds respectively to the solid, dashed and
dotted line. The red, orange, blue and black line indicates QD, CC,
C and GMD respectively. The inset is the uncombined
plot.}\label{fig.5}
\end{figure}

Figures \ref{fig.3} and \ref{fig.5} are plotted to exhibit the
ffects of $B$ and $D$ on various quantities as the external magnetic
field and DM interaction are introduced. One can clearly see that
$B$ plays a destructive role in the manipulation and control of
these correlations from Fig. \ref{fig.3}. But it still deserves to
be studied since the introduction of it is sometime in practical
need such as the nuclear magnetic resonance quantum computing and
the superconducting quantum computing. In particular, the
destructive effect of $B$ can be compensated through adjusting other
tunable system parameters, say $J$ and $D$ for instance. In order to
compare the efficiency of the parameters against the detrimental
$B$, we plot the QD as a function of $B$ and $D$ (or $B$ and $J,$ or
$B$ and $J_z$) in Fig. \ref{fig.4}. The figure shows that the
effects of $J$ and $D$ on QD are exactly equivalent and can tune QD
to the maximal value as long as they are large enough, whereas the
effect of $J_z$ is comparatively weaker (see the lower surface plot
in Fig. \ref{fig.4}) in compensating the detrimental influence of
$B$ on QD. The behavior of other three quantities versus these
system parameters are similar (plots omitted). Furthermore, we
should note that the difference of these quantities becomes larger
as $B$ increases until the critical value, at which these quantities
start at zero and increase as temperature rises (i.e., the value of
any quantity at zero temperature limit transits from nonnull to null
at the critical magnetic field or vice versa). With the further
increase of $B$, such difference vanishes. Moreover, as mentioned
above in Fig. \ref{fig.1}, QD increases with $T$ when $J_z$ is in
the negative region. Such negative-only region can be widened to the
positive one by the inclusion of a strong $B$, and also such
characteristic is awarded to CC and C. Besides, Fig. \ref{fig.5}
shows that DM interaction is constructive for various quantities. We
should note that the difference of QD and CC becomes smaller as $D$
increases and the two curves overlap each other when $D$ is large
enough, while the difference of QD and C becomes larger with $D$.
However, QD and GMD almost overlap each other all the time and the
influence of $D$ on their negligible distinction is very slight. In
addition, $D$ as well as $J$ turn out to be the most efficient
parameters in increasing various correlations as well as the
critical temperature.

\section{The correlations under the intrinsic decoherence}
Now we consider the influence of intrinsic decoherence, proposed by
Milburn \cite{MilburnPRA} with the assumption that a system does not
evolve continuously under unitary transformation for sufficiently
short time steps, on various correlations. The master equation
describing the intrinsic decoherence can be formulated as
\begin{eqnarray}
\label{eq:20}
 \frac{d\rho(t)}{dt} = -i[H, \rho] - \frac{1}{2\gamma}
 [H,[H,\rho(t)]],
\end{eqnarray}
where $\gamma$ is the phase (intrinsic) decoherence rate. The formal
solution of the above equation is given by \cite{CessaPRA}
\begin{eqnarray}
\label{eq:21}
 \rho(t) = \sum_{k=0}^{\infty} \frac{l^k}{k!} M^{k}(t) \rho(0) M^{\dagger k}(t),
\end{eqnarray}
where $\rho(0)$ is the density operator of the initial system and
$M^{k}(t)$ is defined by
\begin{eqnarray}
\label{eq:22}
 M^{k}(t) = H^{k} \exp(-\textrm{i} H t) \exp \bigg(-\frac{t}{2\gamma} H^2
 \bigg).
\end{eqnarray}
Then the time evolution of the density operator for the Heisenberg
$XXZ$ spin system mentioned in the above section can be expressed by
\begin{eqnarray*}
\label{eq:23} \hspace*{6mm}
 \rho(t) = \sum_{mn} \exp \bigg[-\frac{\gamma t}{2} (E_m - E_n)^2 -\textrm{i}(E_m - E_n) t
 \bigg] \qquad
\end{eqnarray*}
\vspace*{-5mm}
\begin{eqnarray}
 \qquad\hspace*{-18mm}\times \langle \psi_m | \rho(0) | \psi_n \rangle | \psi_m \rangle \langle \psi_n
 |,
\end{eqnarray}
where $E_{m,n}$ and $\psi_{m,n}$ are the eigenvalues and the
corresponding eigenvectors of the Hamiltonian respectively.

Firstly, we assume that the initial state of the system is
$|\Psi_{1}(0)\rangle =\frac{1}{\sqrt{2}} \big(| 01 \rangle + | 10
\rangle \big)$ and then $\rho_{1}(0)= | \Psi_{1}(0) \rangle \langle
\Psi_{1}(0)|$. Secondly, we consider another initial state
$|\Psi_{2}(0)\rangle =\frac{1}{\sqrt{2}} \big(| 00 \rangle + | 11
\rangle \big)$. As per the Eqs. (\ref{eq:1}) $\sim$ (\ref{eq:15})
and (\ref{eq:23}), after some algebras, one can obtain the
analytical expressions for the dynamical behavior of various
correlations as
\begin{eqnarray}
\label{eq:24}
\begin{array}{ll}
C_{1}(t) = \mu^{-1} \sqrt{J^{2} + D^{2} e^{-4 \mu^{2}
 \gamma t} \cos(2 \mu t)}, \\
QD_{1}(t) = -\sum\limits_{i=1}^{2} \alpha_{i} \log_{2} \alpha_{i} +
 \sum\limits_{i=3}^{4} \alpha_{i} \log_{2} \alpha_{i}, \\
CC_{1}(t) = -\sum\limits_{i=1}^{2} \alpha_{i} \log_{2} \alpha_{i}, \\
2D_{G1}(t) = \frac{1}{2 \mu^{2}} \big[2J^{2} + D^{2} e^{-4 \mu^{2}
 \gamma t} \big(1+ \cos^{2}(4 \mu t) \big) \big],\\
\end{array}
\end{eqnarray}
for the initial state $|\Psi_{1}(0)\rangle$ and
\begin{eqnarray}
\label{eq:25}
\begin{array}{ll}
C_{2}(t) = e^{-2 B^{2} \gamma t}, \\
QD_{2}(t) = 1+\sum\limits_{i=1}^{2} \beta_{i} \log_{2} \beta_{i}, \\
CC_{2}(t) = 1, \\
2D_{G2}(t) = e^{-4 B^{2} \gamma t},\\
\end{array}
\end{eqnarray}
for the initial state $|\Psi_{2}(0)\rangle$, where $\alpha_{1,2} =
\frac{1}{2} \big[1 \pm \mu^{-1} D e^{-2 \mu^{2} \gamma t} \sin(2\mu
t) \big]$, $\alpha_{3,4} = \frac{1}{2} \big(1 \pm \mu^{-1} e^{-2
\mu^{2} \gamma t} \sqrt{D^{2} + J^{2} e^{2 \mu^{2} \gamma t}}
\big)$, and $\beta_{1,2}= \frac{1}{2} \big(1 \pm e^{-2 B^{2} \gamma
t} \big)$. From the analytical expressions, one can readily see that
the time evolution of these quantities are independent of $J_z$ in
both cases under consideration. Also, they are independent of $B$ in
the case that the initial state is $|\Psi_{1}(0)\rangle$, whereas
independent of $J$ and $D$ in the case that $|\Psi_{2}(0)\rangle$ is
chosen as the initial state.

\begin{figure}[tbp]
\centering
\includegraphics[height=3cm,width=8.7cm]{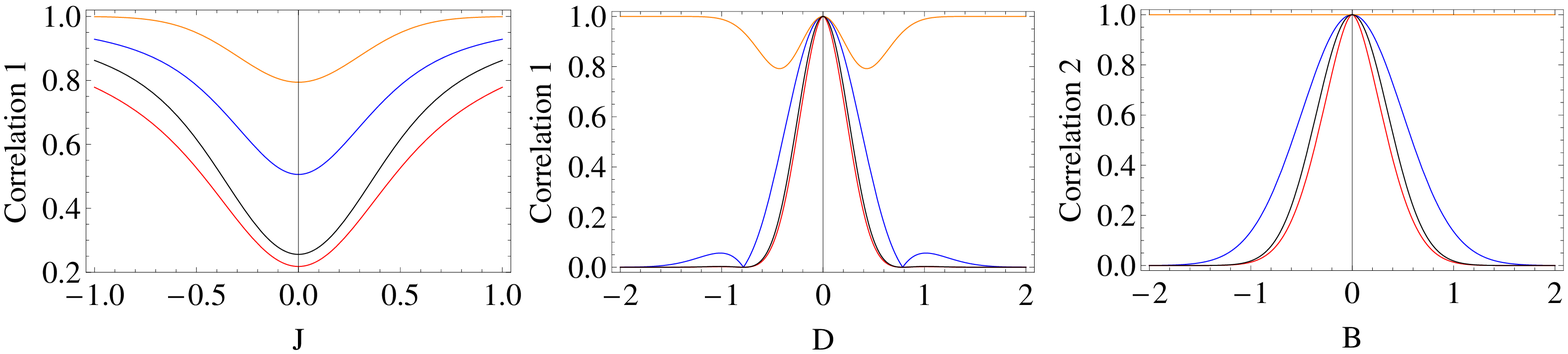}
\caption{(Color online) Time evolution of various correlations are
plotted as a function of $J$ (with $D=0.4$),$D$ (with $J=0$) and $B$
respectively at time $t=1$ with fixed decoherence rate $\gamma=1$,
in which the orange, blue, black and red line corresponds to CC, C,
GMD and QD respectively.}\label{fig.6}
\end{figure}

In what follows, we are dedicated to the numerical results. In Fig.
\ref{fig.6} we plot the time evolution of various quantities versus
$J,$ $D$ and $B$ respectively with other parameters fixed. From the
figure one can see that the effects of $J$ and $B$ on the time
evolution behavior of various quantities are similar to that in
thermal equilibrium except that CC is always maximal and independent
of $B$. However, the effect of $D$ on the dynamics of various
correlations are notably different from its effect in thermal
equilibrium. Note that $D$ is completely detrimental for QD and GMD
and nearly destructive for C except for a process of sudden death
and slight revival, while almost beneficial for CC apart from a
so-called regrowth process that CC decreases to the minimum value
(not zero) as $D$ increases and then increases to the maximal value.

\begin{figure}[tbp]
\centering
\includegraphics[height=5.5cm,width=8cm]{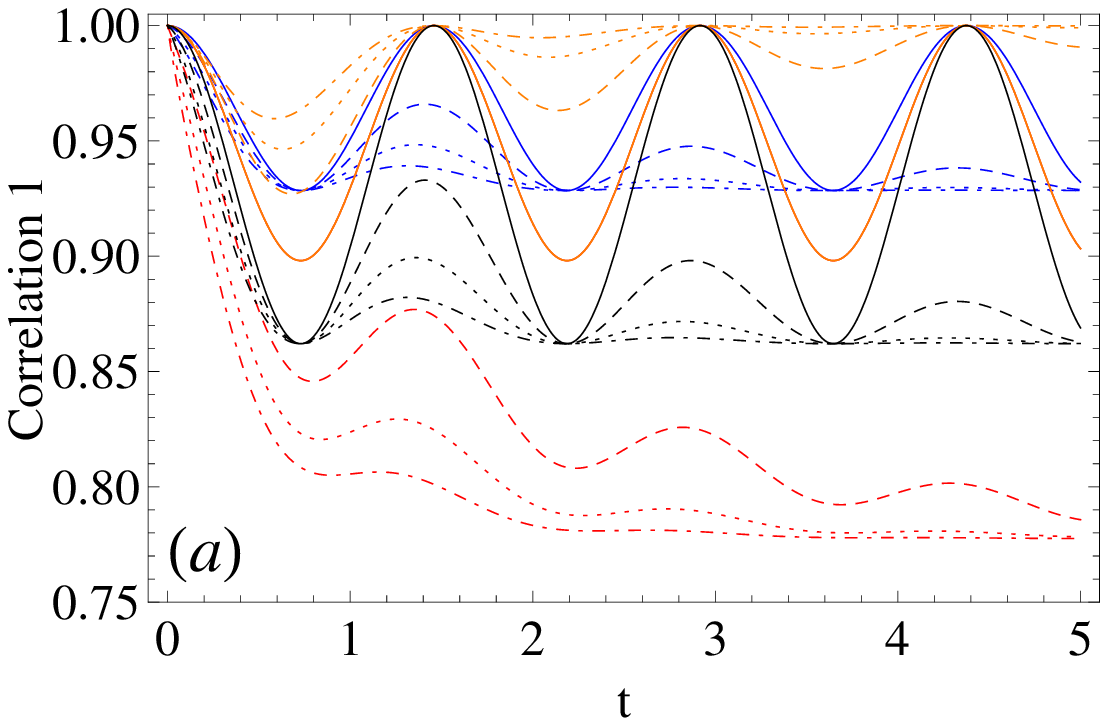}
\includegraphics[height=5.5cm,width=8cm]{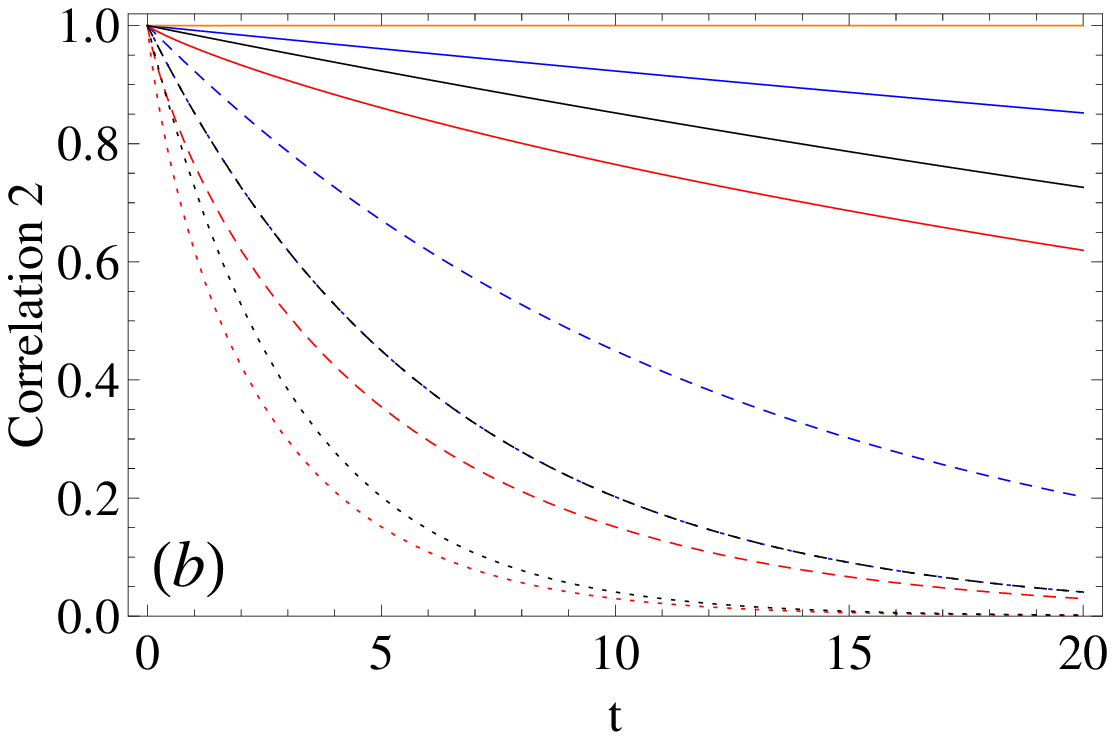}
\caption{(Color online) Time evolution of various correlations are
plotted as a function of time $t$ for different decoherence rate
$\gamma$ with two different initial states, where
$\gamma=0,0.1,0.2,0.3$ corresponds respectively to the solid,
dashed, dotted and dot-dashed line in $(a)$, $\gamma=0.1,1,2$
corresponds to the solid, dashed and dotted line respectively in
$(b)$. The orange, blue, black and red line represents CC, C, GMD
and QD respectively.}\label{fig.7}
\end{figure}

Finally, Fig. \ref{fig.7} is plotted for the two different initial
states in order to observe the effects of pure phase decoherence
rate $\gamma$ on the dynamics of various quantities. Before giving
the numerical analysis, we should clarify that the orange-solid line
and red-solid line in (a) as well as the blue-dotted line and
black-dashed line in (b) are fully overlapped. Figure \ref{fig.7}(a)
depicts that the dynamics of these quantities oscillate with time
$t$ periodically with the same periodicity and the amplitudes of C,
QD and GMD decay gradually to a stable value after a long time
evolution as intrinsic decoherence is taken into account, while that
of CC, on the contrary, is enhanced with the increase of $\gamma$.
The larger $\gamma$ leads to the faster decay (or promotion for CC)
in a short time. Furthermore, when $J=0$ and $\gamma$ is very small,
these quantities undergo the sudden death and revival periodically.
Figure \ref{fig.7}(b) shows that CC is maximal all the time but
other quantities dissipate monotonously and disappear eventually as
time goes to infinity so that we can conclude that CC is robust
against these tunable parameters whereas QD is most sensitive.
Moreover, there is a descending order of CC, C, GMD and QD when the
initial state is $| \Psi_2(0) \rangle$ with a nonnull $B$ (when
$B=0$, these quantities are all maximal).

\section{Conclusion}

In summary, we have investigated various correlations measured by C,
CC, QD and GMD in a two-qubit Heisenberg $XXZ$ spin chain in the
presence of external magnetic field and DM anisotropic antisymmetric
interaction both in thermal equilibrium and under the intrinsic
decoherence cases. We have obtained analytical expressions for these
correlations for both cases and discussed their behaviors following
various system parameters at length. The results show that the
isotropy parameter $J$ plays a constructive role in the manipulation
and control of various correlations and the anisotropy $J_z$ is
considerably crucial for these quantities in thermal equilibrium at
zero temperature limit but ineffective under the consideration of
the intrinsic decoherence. When $J_z<J<-J_z$ ($J_z$ is negative) in
the absence of $B$ and $D$, QD and GMD start at zero and increase as
$T$ rises to a certain value, then decrease, while C is zero and CC
declines from the maximal value with the rise of $T$. For $J=|J_z|$,
all the quantities start at a certain value and decrease with $T$
except for C which is still zero. When $|J|>-J_z$, all quantities
decrease starting from maximum with $T$. Therefore, $J_z=\pm J$ are
the QPT points which are signaled not only by QD and GMD, but also
by CC. As $B$ and $D$ are introduced to the system, the range of
$J_z,$ in which the quantities start at zero and increase with $T$
to a certain value then decrease, are widened to the positive
region. The inclusion of $B$ turns out to be destructive,
nevertheless it still deserves to be studied for its practical
application in some implementations such as the nuclear magnetic
resonance quantum computing and the superconducting quantum
computing. $D$ plays a constructive role and the effects of $J$ and
$D$ on the correlations are exactly equivalent and they turn out to
be the most efficient in compensating the detrimental influence of
$B$. Furthermore, the difference of various quantities becomes
larger with the enhancement of $B$ until the critical point, after
which it minifies. When the intrinsic decoherence is taken into
account, the effect of $J$ and $B$ are similar with that in thermal
equilibrium, but $D$ becomes to be destructive. In addition, the
dynamics of these quantities oscillate with time $t$ when the
initial state is $|\Psi_1(0)\rangle$ and the amplitudes of C, QD and
GMD decay to a stable value after a long time evolution with the
enhancement of $\gamma$, while that of CC, on the contrary, is
enhanced with the increase of $\gamma$. And CC is maximal all the
time but other quantities dissipate degressively and disappear
eventually with $t$ when the initial state is chosen as an
alternative Bell state $|\Psi_2(0)\rangle$. Moreover, there is not a
definite ordering of various quantities in thermal equilibrium,
whereas there is a descending order of CC, C, GMD and QD under the
intrinsic decoherence with a nonnull $B$ when the initial state is
chosen as $|\Psi_2(0) \rangle$.

\begin{acknowledgments}
This work was supported by the National Basic Research Program of
China (973 Program) grant No. G2009CB929300 and the National Natural
Science Foundation of China under Grant No. 60821061. Ahmad Abliz
also acknowledges the Key Subjects of Xinjiang Uygur Autonomous
Region.
\end{acknowledgments}

\end{document}